%Paper: hep-th/9306005
%From: Anatoli KIRILLOV <kirillov@litp.ibp.fr>
%Date: Tue, 1 Jun 1993 14:23:59 -0400
%Date (revised): Wed, 2 Jun 1993 12:46:27 -0400

%=================================================================
% Sergei Fomin, Anatol N. Kirillov  ( Febriary 1993 )
%
% Yang-Baxter equation, symmetric functions and Grothendieck polynomials
%
%=================================================================
%
% Fill in the data (title, authors name, etc.) of your manuscript.
% The places to be filled in occur immediately after the symbols '!',
% hence you can search for '!' one by one.
\font\tenbf=cmbx10
\font\tenrm=cmr10
\font\tenit=cmti10

\font\eightbf=cmbx8
\font\eightrm=cmr8
\font\eightit=cmti8
\font\sevenrm=cmr7 	%
\font\germ=eufm10
\def\eno{\eqalignno}
\def\wt{\widetilde}
\def\sectiontitle#1\par{\vskip0pt plus.1\vsize\penalty-250
 \vskip0pt plus-.1\vsize\bigskip\vskip\parskip
 \message{#1}\leftline{\tenbf#1}\nobreak\vglue 5pt}
\magnification=\magstep1
%\input a4new
%=======================================
\pretolerance=500 \tolerance=1000  \brokenpenalty=5000
\catcode`\@=11
\font\eightrm=cmr8         \font\eighti=cmmi8
\font\eightsy=cmsy8        \font\eightbf=cmbx8
\font\eighttt=cmtt8        \font\eightit=cmti8
\font\eightsl=cmsl8        \font\sixrm=cmr6
\font\sixi=cmmi6           \font\sixsy=cmsy6
\font\sixbf=cmbx6

% Fontes AMS

\font\tengoth=eufm10       \font\tenbboard=msbm10
\font\eightgoth=eufm8      \font\eightbboard=msbm8
\font\sevengoth=eufm7      \font\sevenbboard=msbm7
\font\sixgoth=eufm6        \font\fivegoth=eufm5

% Nouvelles familles pour les maths

\newfam\gothfam           \newfam\bboardfam

\def\tenpoint{%
  \textfont0=\tenrm \scriptfont0=\sevenrm \scriptscriptfont0=\fiverm
  \def\rm{\fam\z@\tenrm}%
  \textfont1=\teni  \scriptfont1=\seveni  \scriptscriptfont1=\fivei
  \def\oldstyle{\fam\@ne\teni}\let\old=\oldstyle
  \textfont2=\tensy \scriptfont2=\sevensy \scriptscriptfont2=\fivesy
  \textfont\gothfam=\tengoth \scriptfont\gothfam=\sevengoth
  \scriptscriptfont\gothfam=\fivegoth
  \def\goth{\fam\gothfam\tengoth}%
  \textfont\bboardfam=\tenbboard \scriptfont\bboardfam=\sevenbboard
  \scriptscriptfont\bboardfam=\sevenbboard
  \def\bb{\fam\bboardfam\tenbboard}%
  \textfont\itfam=\tenit
  \def\it{\fam\itfam\tenit}%
  \textfont\slfam=\tensl
  \def\sl{\fam\slfam\tensl}%
  \textfont\bffam=\tenbf \scriptfont\bffam=\sevenbf
  \scriptscriptfont\bffam=\fivebf
  \def\bf{\fam\bffam\tenbf}%
  \textfont\ttfam=\tentt
  \def\tt{\fam\ttfam\tentt}%
  \abovedisplayskip=12pt plus 3pt minus 9pt
  \belowdisplayskip=\abovedisplayskip
  \abovedisplayshortskip=0pt plus 3pt
  \belowdisplayshortskip=4pt plus 3pt
  \smallskipamount=3pt plus 1pt minus 1pt
  \medskipamount=6pt plus 2pt minus 2pt
  \bigskipamount=12pt plus 4pt minus 4pt
  \normalbaselineskip=12pt
  \setbox\strutbox=\hbox{\vrule height8.5pt depth3.5pt width0pt}%
  \let\bigf@nt=\tenrm       \let\smallf@nt=\sevenrm
  \normalbaselines\rm}

\def\eightpoint{%
  \textfont0=\eightrm \scriptfont0=\sixrm \scriptscriptfont0=\fiverm
  \def\rm{\fam\z@\eightrm}%
  \textfont1=\eighti  \scriptfont1=\sixi  \scriptscriptfont1=\fivei
  \def\oldstyle{\fam\@ne\eighti}\let\old=\oldstyle
  \textfont2=\eightsy \scriptfont2=\sixsy \scriptscriptfont2=\fivesy
  \textfont\gothfam=\eightgoth \scriptfont\gothfam=\sixgoth
  \scriptscriptfont\gothfam=\fivegoth
  \def\goth{\fam\gothfam\eightgoth}%
  \textfont\bboardfam=\eightbboard \scriptfont\bboardfam=\sevenbboard
  \scriptscriptfont\bboardfam=\sevenbboard
  \def\bb{\fam\bboardfam}%
  \textfont\itfam=\eightit
  \def\it{\fam\itfam\eightit}%
  \textfont\slfam=\eightsl
  \def\sl{\fam\slfam\eightsl}%
  \textfont\bffam=\eightbf \scriptfont\bffam=\sixbf
  \scriptscriptfont\bffam=\fivebf
  \def\bf{\fam\bffam\eightbf}%
  \textfont\ttfam=\eighttt
  \def\tt{\fam\ttfam\eighttt}%
  \abovedisplayskip=9pt plus 3pt minus 9pt
  \belowdisplayskip=\abovedisplayskip
  \abovedisplayshortskip=0pt plus 3pt
  \belowdisplayshortskip=3pt plus 3pt
  \smallskipamount=2pt plus 1pt minus 1pt
  \medskipamount=4pt plus 2pt minus 1pt
  \bigskipamount=9pt plus 3pt minus 3pt
  \normalbaselineskip=9pt
  \setbox\strutbox=\hbox{\vrule height7pt depth2pt width0pt}%
  \let\bigf@nt=\eightrm     \let\smallf@nt=\sixrm
  \normalbaselines\rm}

\tenpoint

\def\pc#1{\bigf@nt#1\smallf@nt}         \def\pd#1 {{\pc#1} }

\def\^#1{\if#1i{\accent"5E\i}\else{\accent"5E #1}\fi}
\def\"#1{\if#1i{\accent"7F\i}\else{\accent"7F #1}\fi}

\spaceskip=3pt plus 2pt minus 2pt
 \xspaceskip=\spaceskip

  \catcode`\,=\active  \def,{{\rm \string,}}
%=======================================
\hsize=6.1truein
\vsize=9.2truein
%\hsize=5.0truein
%\vsize=7.8truein
\parindent=15pt
\nopagenumbers
\baselineskip=10pt
\line{
hep-th/9306005\hfil}
%April 1993\hfil}
%\line{\eightrm
%Preliminary version
%Proceedings of the RIMS Research Project 91 on Infinite Analysis
%\hfil}
%\line
%{\hfil NI 92019 }
%{\eightrm $\copyright$\, World Scientific Publishing Company \hfil}
\vglue 5pc
\baselineskip=13pt
\headline{\ifnum\pageno=1\hfil\else%
{\ifodd\pageno\rightheadline \else \leftheadline\fi}\fi}
\def\rightheadline{\hfil\eightit
Yang-Baxter equation, symmetric functions and Grothendieck polynomials
% between dilogarithm function
%! Running title (odd page)
\quad\eightrm\folio}
\def\leftheadline{\eightrm\folio\quad
\eightit
Sergey Fomin and Anatol N. Kirillov
%! Running author (even)
\hfil}
\voffset=2\baselineskip
\centerline{\tenbf
YANG - BAXTER \hskip 0.1cm EQUATION,  \hskip 0.1cm SYMMETRIC \hskip 0.1cm
FUNCTIONS
 }
\centerline{\tenbf
AND \hskip 0.1cm GROTHENDIECK \hskip 0.1cm POLYNOMIALS }
%\hskip 0.1cm CONFORMAL \hskip 0.1 cm
%FIELD \hskip 0.1cm THEORY }
\vglue 24pt
\centerline{\eightrm
SERGEY FOMIN
%! AUTHOR:
%\footnote"$^*$"{\eightrm \baselineskip=10pt
%! present address
% }
}
\baselineskip=12pt
\centerline{\eightit
Department of Mathematics, Massachusetts Institute of Technology,
%! Use the permanent address (University Name, etc.)
}
\baselineskip=10pt
\centerline{\eightit
77 Massachusetts av., Cambridge, MA 02139-4307, U.S.A.
}
\baselineskip=12pt
\centerline{\eightit
and }
\baselineskip=12pt
\centerline{\eightit
Theory of Algorithm Laboratory SPIIRAN,
%! Use the permanent address (University Name, etc.)
}
\baselineskip=10pt
\centerline{\eightit
St.Petersburg, Russia
%! The permanent address (City, State ZIP/Zone, Country)
}
%More than two authors, see the sample prints.
\vglue 20pt
\centerline{\eightrm
ANATOL N. KIRILLOV
%! AUTHOR:
%\footnote"$^*$"{\eightrm \baselineskip=10pt
%! present address
% }
}
\baselineskip=12pt
\centerline{\eightit
L.P.T.H.E., University Paris 6,
%! Use the permanent address (University Name, etc.)
}
\baselineskip=10pt
\centerline{\eightit
4 Place Jussieu, 75252 Paris Cedex 05, France
}
\baselineskip=12pt
\centerline{\eightit
and }
\baselineskip=12pt
\centerline{\eightit
Steklov Mathematical Institute,
%! Use the permanent address (University Name, etc.)
}
\baselineskip=10pt
\centerline{\eightit
Fontanka 27, St.Petersburg, 191011, Russia
%! The permanent address (City, State ZIP/Zone, Country)
}
%More than two authors, see the sample prints.
\vglue 20pt
\centerline{  }
%\centerline{\eightrm Received \quad October 15, 1991}
%\vglue 16pt
%\baselineskip=10pt
%\centerline{\eightrm Revised  {\qquad\qquad} by Publisher)}
%\vglue 20pt
\centerline{\eightrm ABSTRACT}
{\rightskip=1.5pc
\leftskip=1.5pc
\eightrm\parindent=1pc
New development of the theory of Grothendieck polynomials, based on an
exponential solution of the Yang-Baxter equation in the algebra of projectors
are given.
\vglue12pt}
\baselineskip=13pt
\overfullrule=0pt
\font\germ=eufm10
\def\qed{\hfill$\vrule height 2.5mm width 2.5mm depth 0mm$}

{\bf $\S 1$. Introduction.}
\bigbreak

The Yang-Baxter algebra is the algebra generated by operators $h_i(x)$ which
satisfy the following relations (cf. [B], [F], [DWA])
$$\eno{
&h_i(x)h_j(y)=h_j(y)h_i(x),~~{\rm if}~~|i-j|\ge 2,&(1.1)\cr
&h_i(x)h_{i+1}(x+y)h_i(y)=h_{i+1}(y)h_i(x+y)h_{i+1}(x).&(1.2)}
$$
The role the representations of the Yang-Baxter algebra play in the theory of
quantum groups [Dr], [FRT], the theory of exactly solvable models in
statistical mechanics [B], [F], [Ji], low-dimensional topology [DWA], [Jo],
[RT], [KR], the theory of special functions [KR], [VK], and others branchers of
mathematics (see, e.g., the survey [C]) is well-known.

In this paper we continue to study (cf. [FK1], [FK2]) the connections between
the Yang-Baxter algebra and the theory of
symmetric  functions and Schubert and Grothendieck polynomials. In fact, one
can
construct a family of symmetric functions (in general with operator-valued
coefficients) related with any integrable-by-quantum-inverse-scattering-method
model [F]. This class of symmetric functions seems to be very interesting (see
,e.g., [KS]). However, in this paper we want to study ``an inverse problem'',
namely, what kind of ``integrable models'' correspond to the Schur functions
[M1]. Remind that the Schur symmetric functions are the characters of
irreducible, polynomial, finite dimensional representations of the Lie algebra
$su(n)$. Having in mind this ``inverse problem'', let us add to the above
conditions (1.1)-(1.2) an equation
$$h_i(x)h_i(y)=h_i(x+y),~~h_i(0),\eqno (1.3)
$$
thus getting the so-called colored braid relations (CBR), [DWA] (see [KB],
[FS],
[FK1] for examples of their representations). It turns out that, once the
relations (1.1)-(1.3) hold, one can introduce a whole class of symmetric
functions (and
even ``double'', or ``super'' symmetric functions) and respective analogues of
the (double) Schubert (Grothendieck) polynomials [L], [LS], [M2] as well. These
analogues are proved to have many properties of their prototypes; e.g. we
generalize the Cauchy identities and the principal specialization formula (see
also [FK1]).

The simplest solution of the above equations involves the nilCoxeter algebra
of the symmetric group [FS]. Exploring this special case, we constructed in our
previous paper  [FS1] the super-analogues of Stanley's symmetric functions
$F_w$ (see [S]), provided another combinatorial interpretation of Schubert
polynomials $X_w$, and reproved the basic facts concerning $F_w$'s and $X_w$'s.
In this paper we do the same for Grothendieck and stable Grothendieck
polynomials. The Grotendieck polynomials ${\goth G}_w\in S_n$, were introduced
by A.Lascoux and M.-P.Schutzenberger [LS], [L] in their study of Grothendieck
ring of the flag manifold. It is well-known (see e.g. [L]) that Schubert
polynomials $X_w$ corresponds to the homogeneous component of lowest degree in
Grothendieck one ${\goth G}_w$.

The condition (1.2) is one of the forms of the Yang-Baxter equation (YBE);
(1.3) means that we are interested in exponential solutions of the YBE. In
other words, we are interested in quasitriangular $R$ - matrices [D] with
$1$-dimensional auxiliary space. The natural source of such solutions is the
following. Take an associative local algebra with generators $e_i$ (i.e. $e_i$
and $e_j$ commute if $|i-j|\ge 2$) and define $h_i(x)=\exp (xe_i)$. Then (1.1)
and (1.3) are automatically satisfied, and one only needs the YBE (1.2). Let
us sum up our main results of this paper dealing with exponential solutions of
the YBE (see also [FK2]).

{\bf Theorem.} The following statements are equivalent

$i)$ (YBE): $\exp xe_1\cdot\exp (x+y)e_2\cdot\exp ye_1=\exp ye_2\cdot\exp (x+y)
e_1\exp xe_2$;

$ii)$ $\underbrace{[e_1,[e_1,\ldots ,[e_1}_{2n},e_2]\ldots ]]=\underbrace{
[e_2,[e_2,\ldots ,[e_2}_{2n},e_1]\ldots ]]$ for all $n$.

As a corollary we obtain the following examples of exponential solutions of
the YBE:
$$1)~~e_1^2=de_1,~~e_2^2=de_2,~~e_1e_2e_1=e_2e_1e_2\hskip 4.5cm\eqno (1.4)$$
(the nilCoxeter algebra [FS] if $d=0$; the algebra of projectors if $d\ne 0$).
$$2)~~e_1e_2-e_2e_1=1~~{\rm (the \ Heisenberg's\  algebra)}\hskip 4cm\eqno
(1.5)$$
$$3)~~(e_1,(e_1,e_2))=0=(e_2,(e_2,e_1))~~{\rm (the \ algebra}~~U_+(sl(3))
\hskip 2.3cm
\eqno(1.6)
$$
In all these cases $h_i(x)=\exp (xe_i),~~i=1,2$.
$$\eno{
4)~~&(e_1,(e_1,e_2))=0=(e_2,(e_2,e_1))~~{\rm and}~~ h_i(x)=
\exp(x\log (1+e_i)),~~i=1,2,\ \ \ &(1.7)\cr
&{\rm (the~~ generalized~~ Verma~~ relations).}}
$$

The analogues of these examples for the Lie algebras of type $B_2$ and $G_2$
are given in Propositions 2.15 and 2.16.

In sections 3 - 5 we consider in more detail the properties of Grothendieck
polynomials. Among other, we propose a new development of the theory of
Grothendieck polynomials based on an exponential solution (1.4) of YBE. We
propose a new combinatorial rule for description of (stable) Grothendieck
polynomials in terms of 0-1-matrices and give an interpretation of the Coxeter
relations in the symmetric group $S_n$ as some kind of operations (``Coxeter
moves'') on the set of 0-1-matrices (cf. [BB]). We assume to consider the
connections of our construction with that of A. Kohnert [Ko] in a separate
publication. It should be noted that our construction of Grothendieck
polynomials (based on a
study of 0-1-matrices) involvs in consideration also the nonreduced
decompositions of permutations.

It seems a very intriguing problem to give a combinatorial/algebraic
description for (principal) specialization $x_1=1,x_2=1,\ldots $ (resp. $x_1=
1,x_2=q,x_3=q^2,\ldots$) of Grothendieck polynomials (cf. [M2], [FS]). We give
in this paper only a few remarks dealing with this task (see Corollary 3.5).
\medbreak

{\bf Acknowledgement.} The authors are grateful to Alain Lascoux and Richard
Stanley for helpful discussions and useful comments.

\bigbreak

{\bf $\S 2$. The Yang-Baxter equation.}
\bigbreak

Let ${\cal A}$ be an associative algebra with identity 1 over a field $K$ of
zero
characteristic, and let $\{~h_i(x)~|~x\in K,~i=1,2,\ldots \}$ be a family of
elements of ${\cal A}$. We shall study the situation when $h_i(x)$'s satisfy
the
following conditions:
$$\eqalignno{
&h_i(x)h_j(y)=h_j(y)h_i(x),~~{\rm if}~~|i-j|\ge 2,&(2.1)\cr
&h_i(x)h_{i+1}(x+y)h_i(y)=h_{i+1}(y)h_i(x+y)h_{i+1}(x), &(2.2)\cr
&h_i(x)h_i(y)=h_i(x+y),~h_i(0)=1.&(2.3)}
$$
The equation (2.2) is one of the forms of the Yang-Baxter equation (YBE) see
e.g. [DWA]; (2.3)
means that we are interested in exponential solutions of YBE. The most natural
way to construct such solutions is the following. Let $e_1,e_2,\ldots $ be
generators of our algebra ${\cal A}$, assume they satisfy locality condition
$$e_ie_j=e_je_i,~~{\rm if}~~|i-j|\ge 2.\eqno (2.4)
$$
Then let
$$h_i(x)=\exp (xe_i), \eqno (2.5)
$$
we assume that the expression in the right-hand  side  is well-defined in
${\cal A}$.
Then (2.1) and (2.3) are guaranted and we only need to satisfy the YBE (2.2)
which in
this case can be rewritten as
$$\eqalignno{
&\exp (xe_i)\exp (xe_{i+1})\exp (ye_{i+1})\exp (ye_i)=\cr
&=\exp (ye_{i+1})\exp (ye_i)
\exp (xe_i)\exp (xe_{i+1}),& (2.6)}
$$
i.e.
$$\big[~\exp (xe_i)\exp (xe_{i+1}),\exp (ye_{i+1})\exp (ye_i)~\big]=0.
$$
Some examples of solutions are given below.
\medbreak

{\bf Definition 2.1.}  A generalized Hecke algebra
${\cal H}_{a,b}$ is defined to be an associative algebra with generators
$\{e_i : i=1,2,\dots \}$ satisfying
$$
\eqalignno{
&e_ie_j=e_je_i, ~~ |i-j|\ge 2 , \cr
&e_ie_{i+1}e_i=e_{i+1}e_ie_{i+1} , \cr
&e_i^2=ae_i+b\ .}
$$
For example, ${\cal H}_{0,1}$ is the group algebra of the (infinite)
symmetric group; ${\cal H}_{0,0}$ is the corresponding {\it nilCoxeter algebra}
[FS].
\medbreak

{\bf Lemma 2.2.} Let $c\in K$. The elements $h_i(x)\in{\cal H}_{a,b}$
defined by
$$ h_i(x) = 1 + {e^{cx}-1\over a}e_i \eqno(2.7)
$$
satisfy  (2.1)-(2.2).

(This statement is implicit in [R].)

Proof. It is convenient to write $[x]$ instead of ${\displaystyle{e^{cx}-1\over
a}}$.
In this notation, $h_i(x)=1+[x]e_i$. It is easy to check that
$$ [x+y]=[x]+[y]+a[x][y]\ . \eqno (2.8)
$$
Now (cf. (2.2))
$$
\eqalignno{
&(1+[x]e_i)(1+[x+y]e_{i+1})(1+[y]e_i) -
(1+[y]e_{i+1})(1+[x+y]e_i)(1+[x]e_{i+1}) = \cr
&=([x]+[y]-[x+y])(e_i-e_{i+1}) +
[x][y](e_i^2-e_{i+1}^2) = \cr
&-a[x][y](e_i-e_{i+1})+[x][y](ae_i+b-ae_{i+1}-b) = 0\ .}
$$ \qed
%\medbreak

{\bf Corollary 2.3.}  (Case $a=0$.)
Let $p\in K$. The elements $h_i(x)\in{\cal H}_{0,b}$ defined by
$$ h_i(x) = 1+ pxe_i \eqno (2.9) $$
satisfy  (2.1) - (2.2).

Proof. Let in (2.5) $c=ap$ and then tend $a$ to 0.
\qed

In the case $b=p=1$ (the group algebra of the symmetric group)
the example of the previous Corollary is well-known as the so-called
Yang's solution of the Yang-Baxter equation [Y].
\medbreak

{\bf Corollary 2.4.}  (Case $b=0$.)
Let $c\in K$. The elements $h_i(x)\in{\cal H}_{a,0}$ defined by  (2.7)
satisfy  (2.1) - (2.3).

{\it Note.} In fact, one can prove that in this case (2.7)
can be rewritten as
$$ h_i(x) = \exp({c\over a}~xe_i)\ . $$

Proof. We only need to check (2.3) that reduces to (2.6).
\qed

In particular, (2.1) - (2.3) hold for the case $a=b=0$ ([FS], Lemma 3.1).

Now we will construct an exponential solution of the YBE (2.1) - (2.3) with the
values
in the universal enveloping algebra of the Lie algebra of the upper triangular
matrices
with zero main diagonal. This algebra ${\cal A}=U_+({\hbox{\germ g}}l(n))$ can
be defined as
generated by $e_1,\ldots ,e_{n-1}$ satisfying (2.4) and the Serre relations
$$(e_i~,~(e_i,e_{i\pm 1}))=0,\eqno (2.10)
$$
where (~,~) stands for commutator: $(a,b)=ab-ba$. We will show that the
elements
$h_i(x)=\exp (xe_i)$ satisfy (2.1) - (2.3). Hence one can define corresponding
symmetric
functions as well as certain analogues of the Schubert polynomials related to
this
specific solution.
\medbreak

{\bf Theorem 2.5.} Relations (2.4) and (2.10) imply (2.6).
\medbreak

Proof. We give in fact three proofs: ``a computational'' one, based on a
generalization of
Verma's relations in $U_+({\hbox{\germ g}}l(n))$ (see e.g. [V]), another
one based on ``a trick with commutators'',and one based on the
Cambell-Hausdorff formula (see e.g. [Bo]).
\medbreak

{\bf Proposition 2.6.} (Generalization of Verma's relation for $U_+(sl(3)))$.
%\medbreak
$$e_1^n(1+xe_2)^{m+n}e_1^m=(1+xe_2)^me_1^{m+n}(1+xe_2)^n.\eqno (2.11)
$$

Proof. Let us fix $m+n$. If $n=0$, then the identity (2.11) is clear. Now we
use induction
on $n$. Assume (2.11) is valid for $n$, we must prove that (2.11) is still
valid for
$n+1$. Because the algebra $U_+$ does not have the zero divisers it is
sufficient to
prove that
$$(1+xe_2)^{m-1}e_1^{m+n}(1+xe_2)^{n+1}e_1=e_1(1+xe_2)^me_1^{m+n}(1+xe_2)^n.
$$
Now we use the following Lemma.
\medbreak

{\bf Lemma 2.7.} We have
$$\eqalignno{
1)~~&(e_1^N,~e_2)= Ne_1^{ N-1}(e_1,~e_2),\cr
2)~~&(e_1^N,~(e_1,e_2))=0.&(2.12)}
$$

\medbreak
{}From Lemma 2.7 it is easy to deduce that
$$\eqalignno{
&((1+xe_2)^n,e_1)=nx(e_2,e_1)(1+xe_2)^{n-1},\cr
&((1+xe_2)^n,(e_1,e_2))=0.}
$$
Thus it is sufficient to verify the following equality
$$e_1^{m+n+1}(1+xe_2)+e_1^{m+n}(n+1)x(e_2,e_1)=(1+xe_2)e_1^{m+n+1}+mxe_1^{m+n}(e_1,e_2),
$$
which is a direct consequence of (2.12). By the same method one can prove
\medbreak

{\bf Proposition 2.8.} (The generalized Verma relatios for $U_+(sl(3))$).

$$(1+e_1)^n(1+e_2)^{m+n}(1+e_1)^m=(1+e_2)^m(1+e_1)^{m+n}(1+e_2)^n.
$$
\qed

Note that the generalized Verma relations (see Proposition 2.8) are a direct
consequence of ordinary ones
$$e_1^ne_2^{n+m}e_1^m=e_2^me_1^{m+n}e_2^n,~~m,n\in{\bf Z}_+.
$$
This is so because the elements ${\wt e}_i:=e_i+1$, $i+1,2$, also satisfy the
Serre relations. As a corollary of Proposition 2.8 one can easily deduce that
the following elements
$$l_i(x):=\exp (x\log (1+e_i)),~~i=1,2,
$$
give an exponential solution of YBE (see [FK2])
$$l_1(x)l_2(x+y)l_1(y)=l_2(y)l_1(x+y)l_2(x).
$$
As a corollary of Theorem 1, we obtain that elements $a=\log (1+e_1)$ and $b=
\log (1+e_2)$ satisfy the following relations
$$(\underbrace{a,(a,\ldots ,(a}_{2n},b)\ldots ))=(\underbrace{b,(b,\ldots ,
(b}_{2n},a)\ldots ))
$$
for all $n$.

\medbreak

{\bf Corollary 2.9.} $i)$ if $k\le m+n$, then
$$e_1^ne_2^ke_1^m=\pmatrix{m+n\cr k}^{-1}\sum_l\pmatrix{m\cr l}\pmatrix{n\cr
k-l}
e_2^le_1^{m+n}e_2^{k-l},\eqno (2.13~i)
$$
$ii)$ for given $m,n,r,s$ we have an identity
$$\pmatrix{m+n\cr r}\sum_l\pmatrix{n\cr l}\pmatrix{m\cr s-l}e_1^le_2^r
e_1^{s-l}=\pmatrix{m+n\cr s}\sum_l\pmatrix{n\cr l}\pmatrix{m\cr r-l}e_2^{r-l}
e_1^se_2^l.\eqno (2.13~ii)
$$

{\bf Remark 1.} It is easy to see that the generalized Verma relations (see
Proposition 2.8) is equivalent to the identity (2.13~$ii$) and in fact
(2.13~$ii$) follows from (2.13~$i$).

It is well-known (see e.g. [L]) that the following elements of $U_+(sl(3))$
$$\eqalignno{
&e_1^{(n)}e_2^{(p)}e_1^{(m)}~, ~e_2^{(n)}e_1^{(p)}e_2^{(m)}~,~~p>m+n,&(2.14)\cr
&e_1^{(n)}e_2^{(n+m)}e_1^{(m)}=e_2^{(m)}e_1^{(n+m)}e_2^{(n)}~,~~m,n,p\in
{\bf Z}_+,}
$$
where $e^{(n)}:=\displaystyle{{e^n\over n!}}$, give an additive basis (in fact
canonical
basis [L]) of algebra $U_+(sl(3))$.

{}From Corollary (2.9) one can obtain the following rule for decomposing a
monom
$e_1^{(n)}e_2^{(k)}e_1^{(m)}\in U_+(sl(3))$ (resp.
$e_2^{(n)}e_1^{(k)}e_2^{(m)}$) with $k<n+m$ in terms of canonical basis (2.14):
$$\eqalignno{
&e_1^{(n)}e_2^{(k)}e_1^{(m)}=\sum_l\pmatrix{m+n-k\cr m-l}e_2^{(l)}e_1^{(n+m)}
e_2^{(k-l)},\cr
&e_2^{(n)}e_1^{(k)}e_2^{(m)}=\sum_l\pmatrix{m+n-k\cr m-l}e_1^{(l)}e_2^{(m+n)}
e_1^{(k-l)}.}
$$

Now we continue our proof of Theorem 2.5. We must prove the following
\medbreak

{\bf Proposition 2.10.} (Yang-Baxter equation in $U_+(gl(3))$)
$$\big[ \exp (xe_1)\exp (xe_2),\exp (ye_2)\exp (ye_1)\big] =0.\eqno (2.15)
$$

Proof. Let us define
$$T_n=\sum^n_{k=0}\pmatrix{n\cr k}e_1^ke_2^{n-k},~~
S_m=\sum^m_{l=0}\pmatrix{m\cr l}e_2^le_1^{m-l}.
$$
It is clear that
$$\eqalignno{
&T_{n+1}=e_1T_n+T_ne_2,~~S_{m+1}=e_2S_m+S_me_1,\cr
&\exp (xe_1)\exp (xe_2)=\sum{x^n\over n!}T_n,& (2.16)\cr
&\exp (ye_2)\exp (ye_1)=\sum{y^m\over m!}S_m.}
$$
Thus it is sufficient to prove that $[T_n,S_m]=0$ for all $n,m$.

It is clear that
$$\eqalignno{
&T_nS_m=\sum_{k,l}\pmatrix{n\cr k}\pmatrix{m\cr l}e_1^ke_2^{n-k+l}e_1^{m-l},\cr
&S_mT_n=\sum_{k,l}\pmatrix{n\cr k}\pmatrix{m\cr l}e_2^le_1^{m-l+k}e_2^{n-k}.}
$$
We may assume $n\ge m$.
\medbreak

{\bf Lemma 2.11.}
$$\eqalignno{
T_nS_m=S_mT_n=&\sum_{\matrix{k,l\cr k-l<{n-m\over 2}}}\pmatrix{n\cr k}
\pmatrix{m\cr l}e_1^ke_2^{n-k+l}e_1^{m-l}+ &(2.17)\cr
&+\sum_{\matrix{k,l\cr k-l\ge{n-m\over 2}}}\pmatrix{n\cr k}\pmatrix{m\cr l}
e_2^le_1^{m-l+k}e_2^{n-k}.}
$$

Proof. It follows from (2.13) that
$$\eqalignno{
&\sum_{\matrix{k,l\cr n-m\le 2(k-l)}}\pmatrix{n\cr k}\pmatrix{m\cr
l}e_1^ke_2^{n-k+l}e_1^{m-l}=\cr
&=\sum_{k,l,a}{\pmatrix{n\cr k}\pmatrix{m\cr l}\pmatrix{m-l\cr a}
\pmatrix{k\cr n-k+l-a}\over\pmatrix{m-l+k\cr
n-k+l}}~e_2^ae_1^{m-l+k}e_2^{n-k+l-a}=
\cr
&=\sum_{k,l}e_2^le_1^{m-l+k}e_2^{n-k}\sum_a{\pmatrix{n\cr k-l+a}\pmatrix{m\cr
a}
\pmatrix{m-a\cr l}\pmatrix{k-l+a\cr n-k}\over\pmatrix{m+k-l\cr n-k+l}}=\cr
&=\sum_{k,l}\pmatrix{n\cr k}\pmatrix{m\cr l}e_2^le_1^{m-l+k}e_2^{n-k}.}
$$
We used the well-known binomial identity
$$\sum_a\pmatrix{n+k\cr k+a}\pmatrix{m\cr a}\pmatrix{m-a\cr l}\pmatrix{k+a\cr
n-l}=
\pmatrix{m\cr l}\pmatrix{m+k\cr n}\pmatrix{n+k\cr k+l}.
$$

So, we have proved that $T_nS_m$ is equal to the right hand side of (2.17).
Using the same arguments, one can prove that $S_mT_n$ is also equal to the RHS
of (2.17).

\qed

{\it Second proof of Theorem 2.5.}

We need to prove that
$$~(e_1,~(e_1,e_2))=(e_2,~(e_1,e_2))=0~$$
implies
$$[~\exp (xe_1)\exp (xe_2)~,~\exp (ye_2)\exp (ye_1)~].$$
It is sufficient to show that the coefficient $T_n$ of ${\displaystyle{x^n\over
n!}}$ in $\exp (xe_1)\exp (xe_2)$ commutes with the coefficient ~$S_m$~ of ~
${\displaystyle{y^m\over m!}}$~ in $\exp (ye_2)\exp (ye_1)$ (see (2.15)). Let
${\cal L}$ be the algebra generated by $e_1+e_2$ and $(e_1,~e_2)$. We will
prove that $T_n\in{\cal L}$. Then, similary, $S_m\in{\cal L}$ and they commute
because ${\cal L}$ is commutative. Now we have ~$T_{n+1}=e_1T_n+T_ne_2$ (see
(2.16)). So our statement follows from the following Lemma.
\medbreak

{\bf Lemma 2.12.} If $T\in{\cal L}$, then $e_1T+Te_2\in{\cal L}$.
%\medbreak

Proof. Since $e_1T+Te_2=(e_1+e_2)T+(T,e_2)$, we need to prove that $(T,e_2)\in
{\cal L}$. We can assume that $T$ is a momomial in $(e_1,e_2)$ and $e_1+e_2$.
Now take $Te_2$ and move $e_2$ to the left through all the factors; each of
these is either $(e_1+e_2)$ or $(e_1,e_2)$. While moving, we will be getting
on each step an additional term which is either $(e_1+e_2,e_2)$ or
$((e_1,e_2),e_2)$ surrounded by expressions belonging to ${\cal L}$. Since both
$(e_1+e_2,e_2)\in {\cal L}$ and $((e_1,e_2),e_2)\in{\cal L}$ this completes
the proof of Lemma 2.12 and Theorem 2.5.

\qed

{\it Third proof of Theorem 2.5.}

We use the Campbell-Hausdorff formula (CHF) in the free (associative) algebra
with two generators $a$ and $b$:
$$\eno{
&{1\over h}\log (\exp (ha)\cdot\exp (hb))=\cr
&=a+b+{h\over 2}(a,b)+{h^2\over 12}
\{ (a,(a,b))+(b,(b,a))\} +{h^3\over 24}(a,(b,(b,a)))+\cr
&+{h^4\over 144}\{ 9(aabba)+9(bbaab)+4(abbba)+4(baaab)-2(aaaab)-2(bbbba)\}
+O(h^5).}
$$
It is an easy consequence of CHF that if $a$ and $b$ satisfy the Serre
relations
$$(a,(a,b))=0=(b,(b,a))$$
then
%\vskip -1cm
$$\eno{
&\exp a\cdot\exp b=\exp (a+b)\cdot\exp ({1\over 2}(a,b))=\cr
&\exp (a+b+{1\over 2}(a,b))=\exp ({1\over 2}(a,b))\cdot\exp (a+b).}
$$
Consequently,
$$\eno{
\exp xa\cdot\exp xb\cdot\exp yb\cdot\exp ya&=\exp ((x+y)(a+b)+{x^2-y^2\over 2}
(a,b))=\cr
&=\exp yb\cdot\exp ya\cdot\exp xa\cdot\exp xb.}
$$
\qed

{\bf Remark 2.} Consider a $K$ - algebra ${\goth V}$ with two generators
$e_1,~e_2$ and the following system of relations
$$T_nS_m=S_mT_n,~~{\rm for ~~all}~~ m,n\in{\bf Z}_+,\eqno (2.18)
$$
where $T_n$ and $S_m$ are given by (2.15).

It follows from Corollary 2.4 and Theorem 2.5, that there exist the
epimorphisms
$$\eqalignno{
&{\goth V}\longrightarrow U_+(sl(3)),\cr
&{\goth V}\longrightarrow {\cal P}_3=\{~u_1,u_2~|~u_1^2=du_1, u_2^2=du_2,
u_1u_2u_1=u_2u_1u_2~\}.}
$$
It seems an interesting task to rewrite the relations (2.18) in more simple
form. For example, it is easy to see that $T_1S_2=S_2T_1$ iff
$$J_2:=(e_1e_1e_2)-(e_2e_2e_1)=0,
$$
where an expression $(abc)$ is a triple commutator, namely
$$(abc)=a(b,c)-(b,c)a=abc-acb-bca+cba.
$$

{\bf Lemma 2.13.} We have
%If $J_2=0$, then
$$\eqalignno{
i)~~&T_nS_m-S_mT_n\equiv J_{m+n-1}~{\rm mod}~(J_2,\ldots ,J_{n+m-3}),~~~\cr
&{\rm if}~~m+n\equiv 1~({\rm mod}~2),~~m\ge n,\cr
ii)~~&T_nS_m\equiv S_mT_n~{\rm mod}~(J_2,\ldots ,J_{n+m-2})~~{\rm if}~~m+n
\equiv 0~({\rm mod}~2),}
$$
where $J_{2n}:=(\underbrace{e_1,\ldots ,e_1}_{2n}e_2)-(\underbrace{e_2,
\ldots ,e_2}_{2n}e_1)$ is a difference of two $(2n+1)$ - tuple commutators.

Proof. Let us put ${\widetilde J}_n=(e_1+e_2)T_n-T_n(e_1+e_2)$. Note that a
proof of Lemma 2.13 is a consequence of the following statements
$$\eqalignno{
1)~~&{\widetilde J}_n\equiv (-1)^k((\underbrace{e_1,\ldots ,e_1}_k,~e_2),~
T_{n-k})~~{\rm mod}~ ({\widetilde J}_2,\ldots ,{\widetilde J}_{n-1}),\cr
2)~~&{\widetilde J}_n\equiv (-1)^{n-1}((\underbrace{e_1,\ldots ,e_1}_{n-1},~
e_2),~e_1+e_2)~~{\rm mod}~ ({\widetilde J}_2,\ldots ,{\widetilde J}_{n-1}),\cr
3)~~&{\widetilde J}_n\equiv (-1)^{n-k}((\underbrace{e_1,\ldots ,e_1}_k,~e_2,~
(\underbrace{e_1,\ldots ,e_1}_{n-k-1},~e_2))~{\rm mod}~({\widetilde J}_2,\ldots
,{\widetilde J}_{n-1}),\cr
4)~~&{\widetilde J}_{m,n}:=S_mT_n-T_nS_m\equiv\cr
&\equiv (-1)^{k-1}((\underbrace{e_1,\ldots ,e_1}_k,~S_{m-k}),~T_n)~{\rm mod}~
({\widetilde J}_{1,n},\ldots ,{\widetilde J}_{m-1,n}),~~1\le k\le m-1.}
$$It is clear that part 2) follows from 1). Let us prove part 1) using an
induction. Consider $k=1$. We have
$$\eqalignno{
{\widetilde J}_n&=(e_1+e_2)(e_1T_{n-1}+T_{n-1}e_2)-(e_1T_{n-1}+T_{n-1}e_2)
(e_1+e_2)=\cr
&=e_1^2T_{n-1}+e_2e_1T_{n-1}+e_1T_{n-1}e_2+e_2T_{n-1}e_1-\cr
&-e_1T_{n-1}e_1-T_{n-1}
e_2e_1-e_1T_{n-2}e_2-T_{n-1}e_2^2=\cr
&=e_1{\widetilde J}_{n-1}+{\widetilde J}_{n-1}e_2-((e_1,e_2),~T_{n-1}).}
$$
Remind that ${\widetilde J}_{n-1}=(e_1+e_2)T_{n-1}-T_{n-1}(e_1+e_2)$. Now
consider $k\ge 1$. We have
$$\eqalignno{
{\widetilde J}_n&\equiv (-1)^k((\underbrace{e_1\ldots
e_1}_ke_2),~e_1T_{n-k-1})+(-1)^k((\underbrace{e_1\ldots
e_1}_ke_2),~T_{n-k-1}e_2)=\cr
&=(-1)^ke_1((\underbrace{e_1\ldots
e_1}_ke_2),~T_{n-k-1})+(-1)^k((\underbrace{e_1\ldots
e_1}_ke_2),~e_1)T_{n-k-1}+\cr
&+(-1)^kT_{n-k-1}((\underbrace{e_1\ldots
e_1}_ke_2),~e_2)+(-1)^k((\underbrace{e_1\ldots
e_1}_ke_2),~T_{n-k-1})e_2\equiv\cr
&\equiv (-1)^k((\underbrace{e_1\ldots e_1}_ke_2),~e_1)T_{n-k-1}-(-1)^kT_{n-k-1}
((\underbrace{e_1\ldots e_1}_ke_2),~e_1)=\cr
&=(-1)^k(((\underbrace{e_1\ldots e_1}_ke_2),~e_1),T_{n-k-1})=(-1)^{k+1}
((\underbrace{e_1\ldots e_1}_{k+1},e_2),T_{n-k-1}).}
$$
Here all congruences mean those modulo the two-side ideal $({\widetilde
J}_2,\ldots ,{\widetilde J}_{n-1})$.

In oder to check 3) let us put $I_n=(\underbrace{e_1,\ldots ,e_1}_n,~e_2)$.
Then we have
$$\eqalignno{
(-1)^{n-1}{\widetilde J}_n&=(I_{n-1},~e_1+e_2)=(e_1I_{n-2}-I_{n-2}e_1,~e_1+e_2)
\equiv\cr
&\equiv (e_1,e_2)I_{n-2}-I_{n-2}(e_1,e_2)=(I_1,~I_2).}
$$
Now we use an induction. Namely
$$\eqalignno{
(-1)^{n-k}{\widetilde J}_n&\equiv
(I_k,~I_{n-1-k})=(I_k,~e_1I_{n-1-k}-I_{n-1-k}e_1)\equiv\cr
&\equiv ((I_k,~e_1),~I_{n-k-2})=-(I_{k+1},~I_{n-k-2}).}
$$
At last, let us check the statement 4) We have
$$\eqalignno{
{\widetilde J}_{m,n}&=(S_m,T_n)=(e_2S_{m-1}+S_{m-1}e_1T_n)=\cr
&=J_{1,n}S_{m-1}+e_2J_{m-1,n}+J_{m-1,n}e_1+(S_{m-1},~(e_1,T_n))\equiv\cr
&\equiv -((e_1,S_{m-1}),~T_n)~{\rm mod}~({\widetilde J}_{1,n},\ldots ,
{\widetilde J}_{m-1,n}).}
$$
Further, based on an induction assumption, one can find
$$\eqalignno{
{\widetilde J}_{m,n}&\equiv (S_{m-k},~(\underbrace{e_1,\ldots
,e_1}_k,~T_n))=(e_2S_{m-k-1}+S_{m-k-1}e_1,~(\underbrace{e_1,\ldots
,e_1}_k,~T_n))=\cr
&=e_2(S_{m-k-1},~(\underbrace{e_1,\ldots ,e_1}_k,~T_n))+(S_{m-k-1},~
(\underbrace{e_1,\ldots ,e_1}_k,~T_n))e_1+\cr
&+(e_1+e_2,~(\underbrace{e_1,\ldots ,e_1}_k,~T_n))S_{m-k-1}+(S_{m-k-1},~
(\underbrace{e_1,\ldots ,e_1}_{k+1},~T_n))\equiv\cr
&\equiv (S_{m-k-1},~(\underbrace{e_1,\ldots ,e_1}_{k+1},~T_n))\equiv (-1)^k(
(\underbrace{e_1,\ldots ,e_1}_{k+1},~S_{m-k-1}),~T_n).}
$$
Thus all our statements are proved.
\qed

Note that ${\widetilde J}_2=((e_1,e_2),~e_1+e_2)=(e_1e_1e_2)-(e_2e_2e_1)=J_2$.
\medbreak

{\bf Corollary 2.14.} The algebra ${\goth V}$ is isomorphic to that with
generators $u_1$ and $u_2$ subject the relations
$$(\underbrace{u_1,\ldots ,u_1}_{2n},u_2)=(\underbrace{u_2,\ldots ,u_2}_{2n},
u_1),~~n\ge 1.
$$
%Secondly, we have
%$$\eqalignno{
%{\widetilde
%%J}_3&=((e_1e_1e_2),~e_1+e_2)=(e_1(e_1,e_2),~e_1+e_2)-((e_1,e_2)e_1~e_1+e_2)=\cr
%&=e_1((e_1,e_2),~e_1+e_2)-((e_1,e_2),~e_1+e_2)e_1\equiv 0~({\rm mod} J_2).}
%$$
%Using the same arguments one can prove that
%$$\eqalignno{
%&{\widetilde J}_4\equiv ((e_1,e_2),(e_1,e_2),~e_1)~({\rm mod} J_2),\cr
%&{\widetilde J}_5\equiv 0~~{\rm mod} (J_2,J_4).}
%$$
\qed
%\medbreak

{\bf Remark 3.} Theorem 2.5 and Proposition 2.6 may be generalized for the
other Lie algebras of rank two (and so for any finite dimensional semisimple
Lie algebra).
\medbreak

{\bf Proposition 2.15.} Let us denote
$$\eqalignno{
&{\cal
P}(B_2)=\{~e_1,e_2~|~e_1^2=de_1,~e_2^2=de_2,~e_1e_2e_1e_2=e_2e_1e_2e_1~\},\cr
&U_+(B_2)=\{~e_1,e_2~|~(e_1e_1e_1e_2)=0,~(e_2e_2e_1)=0~\}.}
$$
Then we have

i) (Generalized Verma's relations)
$$(1+e_1)^n(1+e_2)^{2n+m}(1+e_1)^{n+m}(1+e_2)^m=(1+e_2)^m(1+e_1)^{n+m}
(1+e_2)^{2n+m}(1+e_1)^n.
$$

ii) (Solutions of CBR). Let us define $h_i(x):=\exp (xe_i),~i=1,2$, where
$e_1$ and $e_2$ are either generators of the algebra ${\cal P}(B_2)$ or
$U_+(B_2)$. Then
$$h_1(x)h_2(2x+y)h_1(x+y)h_2(y)=h_2(y)h_1(x+y)h_2(2x+y)h_1(x).
$$

{\bf Proposition 2.16.} Let us denote
$$\eqalignno{
&{\cal
P}(G_2)=\{~e_1,e_2~|~e_1^2=de_1,~e_2^2=de_2,~e_1e_2e_1e_2e_1e_2=e_2e_1e_2e_1e_2e_1~\},\cr
&U_+(G_2)=\{~e_1,e_2~|~(e_1e_1e_1e_1e_2)=0,~(e_2e_2e_1)=0~\}.}
$$
Then we have

i) (Generalized Verma's relations)
$$\eqalignno{
&(1+e_1)^n(1+e_2)^{3n+m}(1+e_1)^{2n+m}(1+e_2)^{3n+2m}(1+e_1)^{n+m}(1+e_2)^m=\cr
&=(1+e_2)^m(1+e_1)^{n+m}(1+e_2)^{3n+2m}(1+e_1)^{2n+m}(1+e_2)^{3n+m}(1+e_1)^n.}
$$

ii) (Solution of CBR). Let us define $h_i(x):=\exp (xe_i),~i=1,2$,
where $e_1$ and $e_2$ are either the generators of the algebra ${\cal P}(G_2)$
or that of $U_+(G_2)$.

Then
$$\eqalignno{
&h_1(x)h_2(3x+y)h_1(2x+y)h_2(3x+2y)h_1(x+y)h_2(y)=\cr
&=h_2(y)h_1(x+y)h_2(3x+2y)h_1(2x+y)h_2(3x+y)h_1(x).}
$$
\bigbreak
\vskip 0.5cm

{\bf \S 3. Grothendieck and Schubert polynomials.}

\bigbreak

Let $S_n$ denote the symmetric group of permutations of $\{ 1,\cdots ,n \}$.
Let $s_i=(i,i+1),~1\le i\le n-1$, be permutation interchanging $i$ and $i+1$
and leaving all the others elements fixed. For any $w\in S_n$ we denote by
$l(w)$ the length of $w$, i.e. the minimal $l$ such that $w$ can be represented
as
$$w=s_{a_1}s_{a_2}\ldots s_{a_l}$$
for certain $a_1,\ldots ,a_l$; such $a$ sequence $a=(a_1,\ldots ,a_l)$ is
called a reduced decomposition of a permutation $w$ if $l=l(w)$. The set of all
reduced decompositions of $w$ is denoted by $R(w)$. Before giving a
combinatorial definition of the Grothendieck [LS] and the stable Grothendieck
polynomials it is convenient to introduce additional notations. We denote by
$${\cal M}=\{ M=(m_{ij})_{1\le i,j\le n-1}~|~m_{ij}=0~{\rm or}~1\}$$
a set of all $(0,1)$ - matrices of size $(n-1)\times (n-1)$, and let
$${\cal M}^+=\{ ~M\in {\cal M}~|~m_{ij}=0,~~{\rm if}~~1\le j<i\le n-1~\}.$$
a subset of all upper triangular $(0,1)$ - matrices.

For any $M\in {\cal M}$ we define a permutation $s_M\in S_n$ by the following
rule
$$s_M=s_{n-1}^{m_{1,n-1}}s_{n-2}^{m_{1,n-2}}\ldots s_1^{m_{1,1}}\cdot
s_{n-1}^{m_{2,n-1}}\ldots s_1^{m_{2,1}}\ldots s_{n-1}^{m_{n-1,n-1}}\ldots
s_1^{m_{n-1,1}}.\eqno (3.1)
$$
For any $w\in S_n$ let us introduce the subsets
$$\eqalignno{
&{\cal M}(w)=\{ ~M\in{\cal M}~|~s_M=w~\},\cr
&{\cal M}^+(w)=\{ ~M\in{\cal M}^+~|~s_M=w~\}.}
$$
Finally, for any $M\in{\cal M}$ we define
$$\eqalignno{
&x(M)=\prod_{i,j}x_i^{m_{ij}},\cr
&l(M)=\sum_{i,j}m_{ij}.}
$$

{\bf Definition.} Let us fix a parameter $d$ and put
$$\eqalignno{
&{\goth F}_w(x)=\sum_{M\in M(w)}d^{l(M)-l(w)}x(M),\cr
&{\goth G}_w(x)=\sum_{M\in M^+(w)}d^{l(M)-l(w)}x(M). & (3.2)}
$$
\medbreak

{\bf Theorem 3.1.} If $d=-1$, then ${\goth G}_w(x)$ is the Grothendieck
polynomial
corresponding to a permutation $w\in S_n$.
\medbreak

The Grothendieck polynomials $G_w,~~w\in S_n$, were introduced by A.Lascoux and
M.-P.Schutzenberger [LS] in their study of Grothendieck ring of the flag
manifold,
see also [La].

A proof of this Theorem is based on a consideration of generating functions for
polynomials ~~${\goth F}_w(w)$ ~~ and ~~${\goth G}_w(x)$~~. More precisely, let
$K$ be a
commutative ring. We assume $K$ contains various indeterminates used later,
namely $x,x_1,x_2,\ldots ,y,y_1,y_2,\ldots ,q,d$. Define a $K$ - algebra ${\cal
P}_n$ with identity $1$ as follows: ${\cal P}_n$ has generators
$e_1,\ldots ,e_{n-1}$ and relations
%\vskip -0.8cm
$$\eqalignno{
&e_i^2=d~e_i,\cr
&e_ie_j=e_je_i,~~{\rm if}~~|i-j|\ge 2,& (3.3)\cr
&e_ie_{i+1}e_i=e_{i+1}e_ie_{i+1},~~1\le i\le n-2.}
$$
Note that if ~~$d=0$, the algebra ~~${\cal P}_n$
coincides with the nil-Coxeter algebra of the symmetric group (see e.g.
[FS]). If ~~$(a_1,\ldots ,a_l)$~~ is a reduced word, then we will identify a
monomial $e_{a_1}e_{a_2}\ldots e_{a_j}$ in ${\cal P}_n$ with the permutation
$w=s_{a_1}s_{a_2}\ldots s_{a_j}\in S_n$. The relation (3.3) insures that this
notation is well-defined, and ${\cal P}_n$ has $K$ - basis ~~$\{ ~e_w~|~
w\in S_n~\}$.

Define (see [FS])
$$A_i(x)=A_{i,n}(x)=(1+xe_{n-1})(1+xe_{n-2})\ldots (1+xe_i) \eqno (3.4)$$
for $i=1,\ldots ,n-1$. Let
$$\eqalignno{
&{\goth F}(x)=A_1(x_1)A_1(x_2)\ldots A_1(x_{n-1}),\cr
&{\goth G}(x)=A_1(x_1)A_2(x_2)\ldots A_{n-1}(x_{n-1}). & (3.5)}
$$
The following proposition is clear
\medbreak

{\bf Proposition 3.2.} We have
$$\eqalignno{
&{\goth F}(x)=\sum_{w\in S_n}{\goth F}_w(x)\cdot e_w,\cr
&{\goth G}(x)=\sum_{w\in S_n}{\goth G}_w(x)\cdot e_w. & (3.6)}
$$

{\bf Example 1.}  Let us compute ${\goth G}_w$ and ${\goth F}_w$ for
$w=(1432)=s_2s_3s_2$
and $d=-1$. We have
$$\eqalignno{
{\cal M}^+(w)=\Bigg\{ ~~& \matrix{0&1&1\cr &0&1\cr&&0},~~~\matrix{0&1&0\cr
&1&1\cr
 &&0},~~~\matrix{0&1&1\cr&0&0\cr&&1},~~~\matrix{0&0&1\cr &1&0\cr &&1},~~~
\matrix{0&0&0\cr &1&1\cr &&1},\cr
& \cr
&\matrix{0&1&1\cr &1&1\cr &&0},~~~\matrix{0&1&1\cr
&1&0\cr &&1},~~~\matrix{0&1&1\cr &0&1\cr &&1},~~~
\matrix{0&0&1\cr &1&1\cr &&1},~~~\matrix{0&1&0\cr
&1&1\cr &&1},~~~\matrix{0&1&1\cr &1&1\cr &&1} ~~ \Bigg\} . }
$$
Consequently,
$$
{\goth  G}_{1432}=x^2y+xy^2+x^2z+xyz+y^2z-x^2y^2-2x^2yz-2xy^2z+x^2y^2z.
$$
As for stable polynomial ${\goth F}_{1432}$, we have
$$\eqalignno{
{\cal M}(w)={\cal M}(w)^+\cup \Bigg\{ ~& \matrix{0&0&0\cr0&1&0\cr0&1&1},~~~
\matrix{0&1&0\cr0&0&0\cr0&1&1},~~~\matrix{0&1&0\cr0&0&1\cr0&1&0},~~~
\matrix{0&0&0\cr0&1&1\cr 0&1&1},~~~\matrix{0&1&0\cr0&0&1\cr0&1&1},\cr
& \cr
\matrix{0&1&0\cr0&1&0\cr 0&1&1},~~~&\matrix{0&0&1\cr0&1&0\cr0&1&1},~~~
\matrix{0&1&0\cr0&1&1\cr 0&1&0},~~~\matrix{0&1&1\cr0&0&0\cr0&1&1},~~~
\matrix{0&1&1\cr0&0&1\cr 0&1&0},~~~\matrix{0&0&1\cr0&1&1\cr0&1&1},\cr
& \cr
&\matrix{0&1&0\cr0&1&1\cr 0&1&1},~~~\matrix{0&1&1\cr0&0&1\cr0&1&1},~~~
\matrix{0&1&1\cr0&1&0\cr 0&1&1},~~~\matrix{0&1&1\cr0&1&1\cr0&1&0},~~~
\matrix{0&1&1\cr0&1&1\cr 0&1&1}~\Bigg\} .}
$$
Consequently,
$$\eqalignno{
{\goth  F}_{1432}&=x^2y+xy^2+x^2z+xz^2+y^2z+yz^2+2xyz-\cr
&-(x^2y^2+x^2z^2+y^2z^2+3xyz^2+3x^2yz+3xy^2z)+\cr
&+2(x^2y^2z+x^2yz^2+xy^2z^2)-x^2y^2z^2=\cr
&=s_{21}-s_{22}-2s_{211}+2s_{221}-s_{222}.}
$$
Here $s_{\lambda}=s_{\lambda}(x)$ be the Schur function corresponding to
partition
$\lambda$  see e.g. [M1] (in our example $x=(x,y,z)$).

{\bf Remark 1.} It is easy to see that if $d=0$ (the case which was studied
previously in [FS]) then we have: if $M\in{\cal M}(w)$ and
$l(M)>l(w)\Longrightarrow
s_M=0$ (see (3.1) for definition $s_M$).

Now let us consider a matrix $M\in{\cal M}(w)$ with $l(M)=p$. Starting from $M$
we define
two sequences $a=(a_1,\ldots ,a_p)\in{\bf N}^p$ and $b=(b_1,\ldots ,b_p)\in{\bf
N}^p$
using the following rule : let us write the indeces of all non zero entries of
the
matrix $M$ starting from the right to the left and from the top to the bottom:
$(b_1,a_1),\ldots ,(b_p,a_p)$. Then let us put
$$a=a(M) :=(a_1,\ldots , a_p),~~b=b(M) :=(b_1,\ldots ,b_p).
$$
\medbreak

{\bf Example 2.} Consider the following matrix
$$\matrix{0&1&1&0&1\cr0&0&1&0&0\cr0&1&1&1&0\cr1&0&1&1&0\cr0&1&0&0&1}
$$
Then
$$\eqalignno{
(b,a)&=\{ (1,5),~(1,3),~ (12),~ (2,3),~(3,4),~(3,3),~(3,2),~(4,4),~(4,3),~
(4,1),~(5,5),~ (5,2)\},%~{\rm so}~~
\cr
a&=(5~3~2~3~4~3~2~4~3~1~5~2),\cr
b&=(1~1~1~2~3~3~3~4~4~4~5~5).}
$$
\medbreak

{\bf Proposition 3.3.} The correspondence
$$M\to (a(M),b(M))$$
defines a bijection between the following sets
$$\eqalignno{
i)~~&\{ {\cal M}_p(w):=M\in{\cal M}(w)~|~l(M)=p\} ~~{\rm and}\cr
&{\cal D}_p(w):=\left\{ \matrix{&&s_{a}:=s_{a_1}s_{a_2}\ldots s_{a_p}=w,\cr
(a,b)\in{\bf N}^p\times{\bf N}^p&\Bigg\vert &1\le b_1\le\ldots \le b_p\le n-1,
\cr
&&a_i\le a_{i+1}\Longrightarrow b_i<b_{i+1}}\right\} \cr
ii)~~&\{ {\cal M}^+_p(w)=M\in{\cal M}^+(w)~|~l(M)=p\} ~~{\rm and}\cr
&{\cal D}^+_p(w):=\left\{ \matrix{&&\{ j~|~a_j=k\} \le k,\cr
(a,b)\in{\cal D}_p(w)&\Bigg\vert &1\le k\le n-1,\cr
&&b_i\le a_i }\right\} .}
$$
Now let us try to interpret the Coxeter relations in the symmetric group $S_n$
in terms of $(0,1)$-matrices (compare with [BB]). For this purpose let us
define some operations on
the sets ${\cal M}(w)$ and ${\cal M}^+(w),~w\in S_n$. Given $M=(m_{ij})\in
{\cal M}^+(w)$, assume that there exist $1\le p\le q$ and $k\ge 0,~q+k\le n-1$,
such that
$$\eqalignno{
&m_{p,q+1}=\ldots =m_{p,q+k}=1,~m_{p,q+k+1}=0,\cr
&m_{p-1,q-1}=0,~m_{p-1,q}=\ldots =m_{p-1,q+k-1}=1,\cr
&m_{p,q}+m_{p-1,q+k}\ne 0.}
$$
Here we assume that $m_{ij}=0$ if $(i,j)\not\in[1,n-1]\times[1,n-1]$. Let us
define the new matrices ${\widetilde M}^{(\epsilon)}:=({\widetilde m}_{ij}^
{(\epsilon)}),~\epsilon =0,\pm 1$, by the following rules
$$\eqalignno{
i)~~&{\rm if}~~ m_{p,q}+m_{p-1,q+k}=1,~~{\rm then}~~{\widetilde m}_{ij}^{(0)}=
m_{ij}, ~~{\rm if}~~(i,j)\ne (p,q),~~(p-1,q+k),\cr
&{\rm and}~~{\widetilde m}_{p,q}^{(0)}=m_{p-1,q+k},~~
{\widetilde m}_{p-1,q+k}^{(0)}=m_{pq}.\cr
ii)~~&{\rm if}~~ m_{p,q}=1~~{\rm and}~~m_{p-1,q+k}=1,~~{\rm then}\cr
&{\widetilde m}_{ij}^+=m_{ij},~~{\rm if}~~(i,j)\ne (p,q)~~{\rm
and}~~{\widetilde m}_{p,q}^+
=0,\cr
&{\widetilde m}_{ij}^-=m_{ij},~~{\rm if}~~(i,j)\ne (p-1,q+k)~~{\rm
and}~~{\widetilde m}_{p-1,q+k}^-=0.}
$$
We define the operations $\nabla^{\epsilon}_{(p,q,k)}$ on the set of
$(0,1)$-matrices as
$$\nabla^{(\epsilon)}_{(p,q,k)}(M):=({\widetilde m}_{ij}^{(\epsilon)}).
$$

{\bf Proposition 3.4.} i) If $M\in{\cal M}_p^+(w)$, then
$$\nabla_{(p,q,k)}^{(0)}(M)\in{\cal M}_p^+(w).
$$
ii) if $M\in{\cal M}_p^+(w)$ and the operation $\nabla_{(p,q,k)}^{+(-)}$ is
applicable to $M$, then
$$\nabla_{(p,q,k)}^{+(-)}(M)\in{\cal M}_{p-1}(w).
$$

{\bf Remark 2.} It is easy to check by induction, that if $d=-1$, then
$$\eqalignno{
&{\goth F}(1,\ldots ,1)={\goth G}(1,\ldots ,1)=\sum_{w\in S_n}e_w,~~{\rm
i.e.}\cr
&{\goth F}_w(1,\ldots ,1)={\goth  G}_w(1,\ldots ,1)=1~~{\rm for ~~all} ~~
w\in S_n.}
$$
Moreover, if $d=1$, then ${\goth G}_w(1,\ldots ,1)$ is an odd integer for all
$w\in S_n$.

For general $d$ a computation of polynomials $p_w(d):={\goth G}_w(1,\ldots ,1)$
seems
to be a very intriguing problem. Here we give only a few remarks concerning
with this
task. First of all, it is easy to see from Proposition 3.2 that if $w\in S_n$
is a
dominant permutation of shape $\lambda$ (see e.g. [M2], chapter IV), then
$${\goth G}_w(x)=X_w(x)=x^{\lambda},$$
where$X_w(x)$ is the Schubert polynomial corresponding to $w$. So, for a
dominant permutation $w\in S_n$ we have
$${\goth G}_w(1,\ldots ,1)=X_w(1,\ldots ,1)=1.
$$
Conversely, if ${\goth G}_w(1,\ldots ,1)=1$ (for general $d$), then $w$ is a
dominant
permutation. Remind that the number of the dominant permutations in the
symmetric group
$S_n$ is equal to the Catalan number ${\displaystyle C_n={(2n)!\over
n!(n+1)!}}$.
\medbreak

{\bf Lemma 3.5.} In the $K$ - algebra ${\cal P}_3$ we have an identity
$$\eqalignno{
(1+e_1+e_2+e_2e_1)^n&(1+e_2)=1+{(1+d)^n-1\over d}e_1+{(1+d)^{n+1}-1\over
d}e_2+\cr
&+\left({n(1+d)^{n+1}-(n+1)(1+d)^n+1\over d^2}\right)(e_1e_2+e_2e_1)+\cr
&+\left({(1+d)^{2n+1}-1-d(2n+1)(1+d)^n\over d^3}\right) e_1e_2e_1.}
$$
\medbreak

{\bf Corollary 3.6.} We have
$$\eqalignno{
i)~~& p_k:=p_{s_k}(d)={(1+d)^k-1\over d}.\cr
ii)~~&p_{(k+1,k)}:=p_{s_{k+1}s_k}(d)=p_{s_ks_{k+1}}(d)={k(1+d)^{k+1}-(k+1)(1+d)^k+1
\over d^2}=\cr
&={1\over d}(kp_{k+1}-(k+1)p_k).}
$$

More generally,
$$\eqalignno{
&p_{s_ks_{k+1}\ldots s_{k+m-1}}(d)=\cr
&={1\over d^{m-1}}\pmatrix{m+k-1\cr m}\left\{\sum^m_{j=1}(-1)^{j-1}{j\over
k+m-j}\pmatrix{m\cr j}p_{k+m-j}(d)\right\}=\cr
&=\sum_{j=0}^{k-1}\pmatrix{m+k-1\cr m+j}\pmatrix{m+j-1\cr j}d^j.}
$$
In terms of generating function
$$\sum_{k\ge 1}p_{(k,k+1,\ldots ,k+m-1)}(d)z^{k-1}={1\over (1-z)(1-z(1+d))^m}.
$$
$$\eqalignno{
iii)~~p_{s_ks_{k+1}s_k}(d)&={(1+d)^{2k+1}-1-(2k+1)d(1+d)^k\over
d^3}=~~~~~~~~~~~~~~~~~~~~~~~~~\cr
&={1\over d}(p_kp_{k+1}-2p_{(k+1,k)}).}
$$

{\bf Remark 3.} Note that the number of all $(n-1)$ by $(n-1)$ upper triangular
$(0,1)$ - matrices is equal to $2^N$, where $N={\displaystyle{n(n-1)\over 2}}$.
This number is also equal to that of all Gelfand-Tsetlin patterns (GT-patterns)
(see e.g. [KB]) with the highest weight
$\rho_n=(n-1,n-2,\ldots 1,0)$, or equivalently, to the number of all
(semi)standard Young tableaux of the shape $\rho_n$ filled by the numbers
$1,2,\ldots ,n$. Thus it is natural to ask: does there exist a bijection
$$\theta : {\cal M}_n^+\to GT(\rho_n),
$$
such that $\theta({\cal M}_n^+)$ lies in $SGT(\rho_n)$~?

Here ${\cal M}^+_{n,red}=\{ M\in{\cal M}_n^+~|~l(M)=l(s_M)\} $
be the set of all
matrices $M\in{\cal M}^+_n$, having the reduced-decomposition-permutation
$s_M$, and $SGT(\rho_n)$ be the set of all strict Gelfand-Tsetlin patterns.
Remind that a GT-pattern is called strict [T] if each row of it is a strictly
decreasing sequence. It is known ([MRR] or [T]) that there exist a bijection
between $SGT(\rho _n)$ and ${\rm Alt}_n$, the set of $n$ by $n$ alternating
sign matrices. Remind [MRR] that $n$ by $n$ matrix $A=(a_{ij})$
is called an alternating sign matrix if it satisfies the following conditions
$$\eqalignno{
i)~~&a_{ij}\in\{ -1,0,1\} ~~{\rm for ~~all}~~1\le i,j\le n,\cr
ii)~~&\sum_{i=1}^ka_{ij}=0~~{\rm or}~~1~~{\rm and}~~\sum^k_{j=1}a_{ij}=0~~{\rm
or}~~1~~{\rm for~~any}~~1\le k\le n,\cr
iii)~~&\sum^n_{i=1}a_{ij}=\sum^n_{j=1}a_{ij}=1.}
$$
It is the well-known conjecture [MRR] (and now a theorem, see [Z]) that
$$\#|{\rm Alt}_n|=\prod_{i=0}^{n-1}{(3i+1)!\over (n+i)!}.
$$
\bigbreak
%\vskip 1.0cm

{\bf \S 4. Proof of the Theorem 3.1.}
\bigbreak

First of all we study the properties of elements $A_i(x)$ (see (3.4)).
\medbreak

{\bf Lemma 4.1.} Let $e_1,\ldots ,e_{n-1}$ be the generators of algebra ${\cal
P}_n$ (see
(3.3)) and $h_i(x):=1+xe_i$. Then
$$\eqalignno{
&1)~~h_i(x)h_j(y)=h_j(y)h_i(x),~~{\rm if }~~|j-i|\ge 2,& (4.1)\cr
&2)~~h_i(x)h_{i+1}(x\oplus y)h_i(y)=h_{i+1}(y)h_i(x\oplus y)h_{i+1}(x),~~
1\le i\le n-2,\cr
&3)~~h_i(x)h_i(y)=h_i(x\oplus y),}
$$
where $x\oplus y :=x+y+dxy$.

See e.g. [FK] Lemma 2.2.
\medbreak

{\bf Corollary 4.2.} For all $1\le i\le n-1$, we have
$$A_i(x)A_i(y)=A_i(y)A_i(x).
$$
A proof is the same as in [FS] or [FK].
\medbreak

{\bf Lemma 4.3.} Let us put ${\overline e}=e_i-d$. Then
$$A_i(x)A_{i+1}(y){\overline e}_i=A_i(y)A_{i+1}(x){\overline e}_i.
$$
Proof (see [FS]). This identity means that the left-hand side is symmetric in
$x$ and
$y$. Write
$$\eqalignno{
& A_i(x)A_{i+1}(y){\overline e}_i=A_{i+1}(x)(1+xe_i)A_{i+1}(y){\overline
e}_i=\cr
&=A_{i+1}(x)A_{i+1}(y){\overline
e}_i+xA_{i+1}(x)e_iA_{i+2}(y)(1+ye_{i+1}){\overline e}_i.}
$$
and note that the first summand is symmetric by Corollary 4.2. Now
$$\eqalignno{
&xA_{i+1}(x)A_{i+2}(y)e_i(1+ye_{i+1}){\overline
e}_i=xA_{i+1}(x)A_{i+2}(y)ye_ie_{i+1}
{\overline e}_i=\cr
&=xA_{i+1}(x)A_{i+2}(y)y{\overline
e}_{i+1}e_ie_{i+1}=xA_{i+1}(x)A_{i+2}(y)y(1+ye_{i+1}){\overline e}_{i+1}e_i
e_{i+1}=\cr
&=xyA_{i+1}(x)A_{i+1}(y){\overline e}_{i+1}}
$$
which is also symmetric. Here we use the following relations between $e_i$ and
${\overline e}_j$:
$$e_i{\overline e}_i=0,~~e_ie_{i+1}{\overline e}_i={\overline
e}_{i+1}e_ie_{i+1}.$$
\medbreak

{\bf Lemma 4.4.} We have
$$(1+dy)A_i(x)A_{i+1}(y)-(1+dx)A_i(y)A_{i+1}(x)=(x-y)A_i(x)A_{i+1}(y){\overline
e}_i$$

Proof (see [FS]).
$$\eqalignno{
&(1+dy)A_i(x)A_{i+1}(y)+yA_i(x)A_{i+1}(y){\overline
e}_i=A_i(x)A_{i+1}(y)(1+dy+y
{\overline e}_i)=\cr
&=A_i(x)A_{i+1}(y)(1+ye_i)=A_i(x)A_i(y)=A_i(y)A_i(x)=~~~~~({\rm Corollary~~
4.2})\cr
&=A_i(y)A_{i+1}(x)(1+xe_i)=A_i(y)A_{i+1}(x)(1+dx+x{\overline e}_i)=\cr
&=(1+dx)A_i(y)A_{i+1}(x)+xA_i(y)A_{i+1}(x){\overline e}_i=\cr
&=(1+dx)A_i(y)A_{i+1}(x)+xA_i(x)A_{i+1}(y){\overline e}_i~~~({\rm
Lemma~~4.3}).}
$$
\medbreak

{\bf Definition.} (Grothendieck polynomials [LS], [La]). Define the divided
difference
operator ${\cal \pi}_{xy}$ acting in $K$ by
$${\cal\pi}_{xy}f(x,y)={(1+dy)f(x,y)-(1+dx)f(y,x)\over x-y}$$
and let us put ${\cal\pi}_i={\cal\pi}_{x_i,x_{i+1}}$. Grothendieck polynomials
can be
defined recursively by

(i) $~G_{w_0}(x)=x_1^{n-1}x_2^{n-2}\ldots x_{n-1}$, for $w_0=(n,n-1,\ldots
,1)$,

(ii) $~G_w(x)=\pi_iG_{ws_i}(x)$ whenever $w\in S_n,~~l(ws_i)=l(w)+1$.

\medbreak

{\bf Lemma 4.5.} ${\cal\pi}_i{\goth G}(x)={\goth G}(x)\cdot{\overline e}_i$
\medbreak

Proof. Since
$$ {\cal\pi}_i{\goth
G}(x)=A_1(x_1)\cdots{\cal\pi}_i(A_i(x_i)A_{i+1}(x_{i+1}))\ldots
$$
and
$${\goth G}(x){\overline e}_i=A_1(x_1)\ldots
(A_i(x_i)A_{i+1}(x_{i+1}){\overline e}_i)
A_{i+2}(x_{i+2})\ldots ,$$
the statement of Lemma follows from
$${\cal\pi}_i(A_i(x_i)A_{i+1}(x_{i+1}))=A_i(x_i)A_{i+1}(x_{i+1}){\overline
e}_i,
$$
which is exactly Lemma 4.4.

Proof of Theorem 3.1. Take $d=-1$. We need to prove that
$$\eqalignno{
&<{\goth G}(x),~w_0>=x_1^{n-1}x_2^{n-2}\ldots x_{n-1}~~{\rm and}&(4.2)\cr
&\pi_i<{\goth G}(x),~ws_i>=<{\goth G}(x),~w>,~~{\rm if}~~l(ws_i)=l(w)+1.&(4.3)}
$$
The equality (4.2) follows from (3.4) and (3.5): to obtain from 1 the
permutation $w_0$ one should take $x_ie_j$ from each factor in the expansion of
${\goth G}(x)$; (4.3) follows from Lemma 4.5.

\qed

\bigbreak

{\bf \S 5. Cauchy formula and double Grothendieck polynomials.}
\bigbreak

Define
$$\eqalignno{
&{\widetilde A}_i(y)=h_i(y)\ldots h_{n-1}(y),\cr
&{\widetilde{\goth G}}={\widetilde A}_{n-1}(y_{n-1})\ldots {\widetilde A}_1
(y_1).}
$$
\medbreak

{\bf Lemma 5.1.} $A_i(x)$ and ${\widetilde A}_i(y)$ commute.
\medbreak

Proof. Descending induction on $i$, using the colored braid relations (4.1).
(Compere
with a proof of Lemma 4.1 in [FS]).
\qed
\medbreak

{\bf Lemma 5.2.}
$${\widetilde A}_{n-1}(y_{n-1})\ldots {\widetilde
A}_i(y_i)A_i(x)=h_{n-1}(y_{n-1}
\oplus x)\ldots h_i(y_i\oplus x){\widetilde A}_{n-1}(y_{n-2})\ldots {\widetilde
A}_{i+1}(y_i).
$$
A proof is the same as in [FS]. Let us remind that
$$x\oplus y=x+y+dxy,~~x\ominus y={x-y\over 1+dy}.$$
%\medbreak

{\bf Lemma 5.3.}
$${\widetilde{\goth G}}(y){\goth G}(x)=A_1(x_1~|~y_1,\ldots
,y_{n-1})A_2(x_2~|~y_2,\ldots ,y_{n-1})\ldots A_{n-1}(x_{n-1}~|~y_{n-1}),\eqno
(5.1)
$$
where $A_i(x~|~y_i,\ldots ,y_{n-1})=h_{n-1}(x\oplus y_{n-1})h_{n-2}(x\oplus
y_{n-2})\ldots h_i(x\oplus y_i)$.

Proof. Repeatedly apply Lemma 5.2.

\qed

Let us remark that the proofs of Lemmas 5.1 - 5.3 are based on the colored
braid
relations (4.1) only.
\medbreak

{\bf Corollary 5.4.} (Cauchy formula for Grothendieck polynomials)
$$<{\widetilde{\goth G}}(y){\goth G}(x),w_0>=\prod_{i+j\le n}(x_i+y_j+dx_iy_j),
$$
where $w_0=(n,n-1,\ldots ,1)$.

Proof. The right-hand side of (5.1) contain exactly $\pmatrix{n\cr 2}$ factors.
Thus to
obtain $w_0$ from 1 one should take $(x_i\oplus y_j)e_{i+j-1}$ from each
factor.
\qed
\medbreak

{\bf Lemma 5.5.} Let $w\in S_n$. The polynomial
$<{\widetilde{\goth G}}(y){\goth G}(x),w>$ is the double Grothendieck
polynomial
$G_w(x,-y)$ of Lascoux-Schutzenberger [LS],[La].

Proof. (see [FS]). Descending induction on $l(w)$. The basis $(w=w_0)$ is
exactly Corollary 5.4. The induction step follows immediatly from Lemma 4.5
with $d=-1$.

\qed

\medbreak

{\bf Lemma 5.6.}
$${\goth G}\left( {e^{dt}-1\over d},\ldots ,{ e^{dt}-1\over d}\right)=\exp
(t\cdot f),
$$
where $f=e_1+2e_2+3e_3+\ldots$.

Proof. In the case $y_1=y_2=\ldots =y_{n-1}=y$ one can check that
$${\widetilde{\goth G}}(y)={\goth G}(y) =A_1(y)A_2(y)\ldots A_{n-1}(y).
$$
Hence Lemma 5.3 gives
$${\goth G}(y){\goth G}(x)=A_1(x\oplus y)A_2(x\oplus y)\ldots A_{n-1}(x\oplus
y)={\goth G}(x\oplus y).\eqno (5.2)
$$
{}From (5.2) it follows that
$${\goth G}\left( {e^{dt}-1\over d},\ldots ,{e^{dt}-1\over
d}\right)=\exp(t\cdot f).
\eqno (5.3)
$$
Differentiation of (5.3) with respect to $t$ shows that
$$f={d\over dt}{\goth G}(t,\ldots ,t)\mid_{t=0}.
$$
{}From definitions (3.4) and (3.5) and the Leibniz rule we obtain
$f=e_1+2e_2+3e_3+\ldots$,
as desired.
\qed
\medbreak

{\bf Corollary 5.7.}
$${\goth G}(1,\ldots ,1)=\exp\left({\log (1+d)\over d}f\right).
$$
\medbreak

{\bf Lemma 5.8.}
$${\goth G}(1,q,\ldots
,q^{n-2})=\prod^0_{k=\infty}\prod^1_{j=n-1}h_j\left({q^k-q^{k+j}
\over 1+dq^{k+j}}\right).
$$
where in the (non-commutative) products the factors are multiplied in
decreasing order (with respect to $k$ and $j$).

Proof. (see [FS]). Let us prove by induction that
$${\goth G}(q^i,q^{i+1},\ldots ,q^{n+i-2})={\goth G}(q^{i+1},\ldots ,q^{n+i-1})
\prod^1_{j=n-1}h_j(q^i\ominus q^{i+j}),\eqno (5.5)
$$
where in the (non commutative) product the factors are multiplied in decreasing
order (with respect to $j$) and
$$x\ominus y={x-y\over 1+dy}.
$$
Repeatedly using (5.5), one can obtain (5.4). Now
$$\eqalignno{
&{\goth G}(q^{i+1},\ldots ,q^{n+i-1})h_{n-1}(q^i\ominus q^{n+i-1})\ldots
h_1(q^i\ominus q^{i+1})=\cr
&=A_1(q^{i+1})\ldots A_{n-2}(q^{n+i-2})A_{n-1}(q^{n+1-1})h_{n-1}(q^i\ominus
q^{n+i-1})\ldots h_1(q^i\ominus q^{i+1})=\cr
&=A_2(q^{i+1})h_1(q^{i+1})A_3(q^{i+2})h_3(q^{i+2})\ldots A_{n-1}(q^{n+i-2})
h_{n-2}(q^{n+i-2})\cdot\cr
&~~~\cdot h_{n-1}(q^{n+i-1})h_{n-1}(q^i\ominus q^{n+i-1})h_{n-2}(q^i\ominus
q^{n+i-2})\ldots h_1(q^i\ominus q^{i+1})=\cr
&=A_2(q^{i+1})A_3(q^{i+2})\ldots A_{n-1}(q^{n+i-2})h_1(q^{i+1})h_2(q^{i+2})
\ldots h_{n-3}(q^{n+i-3})\cdot\cr
&~~~\cdot h_{n-2}(q^{n+i-2})h_{n-1}(q^i)h_{n-2}(q^i\ominus
q^{n+i-2})\ldots h_{n-1}(q^i\ominus q^{i+1})=\cr
&~~~{\rm (by\  repeated\ use\ of\ Lemma\ 4.1)}\cr
&=A_2(q^{i+1})\ldots A_{n-1}(q^{n+i-2})h_{n-1}(q^i\ominus q^{n+i-2})\ldots
h_2(q^i\ominus q^{i+1})\cdot\cr
&~~~\cdot h_1(q^i)\ldots h_{n-1}(q^{n+i-2})=\cr
&=A_2(q^i)\ldots A_{n-1}(q^{n+i-3})h_1(q^i)\ldots h_{n-1}(q^{n+1-2})=~~
{\rm (by\ the\ induction\ hypothesis)}\cr
&=A_1(q^i)\ldots A_{n-2}(q^{n+i-3})h_{n-1}(q^{n+i-2})={\goth G}(q^i,\ldots ,
q^{n+i-2}).}
$$
\qed

\medbreak
{\bf Remark 1.} Define
$$\eqalignno{
&{\hat A}_i(x)=(1+xe_i)(1+xe_{i-1})\ldots (1+xe_1),\cr
&{\hat{\goth G}}(x)={\hat A}_1(x_1)\ldots {\hat A}(x_{n-1})=\sum_{w\in
S_n}{\hat{\goth G}}_w(x)e_w.}
$$

{\bf Lemma 5.9.} We have
$${\hat{\goth G}}_w(x_1,\ldots ,x_{n-1})={\goth
G}_{w_0w^{-1}w_0}(x_{n-1},\ldots ,x_1),~~w\in S_n.\eqno (5.6)
$$

{\bf Corollary 5.10.} (Expression of the stable Grothendieck polynomials in
terms of Grothendieck polynomials)
$${\goth F}_w(x_1,\ldots ,x_{n-1})=\sum_{w=u\cdot v}{\goth G}_u(x_1,\ldots ,
x_{n-1}){\goth G}_{w_0v^{-1}w_0}(x_{n-1},\ldots ,x_2).\eqno (5.7)
$$

Proof. It is easy to check by induction, that
$${\goth F}(x_1,\ldots ,x_{n-1})={\goth G}(x_1,\ldots ,x_{n-1}){\hat{\goth G}}
(x_2,\ldots ,x_{n-1}).
$$
Thus (5.7) follows from (5.6).
\qed

\vfill\eject
\vskip 1cm
{\bf References.}
\bigbreak

\item{[B]} R.J. Baxter, Exactly solved models in statistical mechanics,
Academic Press, 1982.
\item{[BB]} N. Bergeron, S.C. Billey, $RC$ - graphs and Schubert polynomials,
Preprint, 1992.
\item{[BJS]} S.C. Billey, W. Jockush and R.P. Stanley, Some combinatorial
properties of Schubert polynomials, manuscript, MIT, 1992.
\item{[Bo]} N. Bourbaki, Groupes et algebras de Lie, Paris: Hermann, 1972.
\item{[C]} P. Cartier, Developments recent sur les groupes de tresses.
Applications a la topologie et a l'algebre. Seminaire Bourbaki, vol. 1989/90,
Asterisque 189-90 (1990), Exp. No. 716, 17-47.
\item{[Ch]} I. Cherednik, Notes on affine Hecke algebras. I,
Max-Planck-Institut Preprint MPI/91-14, 1991.
\item{[D]} V. Drinfeld, Quantum groups, Proc. Int. Congr. Math., vol. 1,
798-820, Berkeley, 1987.
\item{[F]} L.D. Faddeev, Integrable models in $1+1$ dimensional quantum field
theory, in Les Houches Lectures 1982, Elsevier, Amsterdam, 1984.
\item{[FK1]} S. Fomin, A.N. Kirillov, The Yang-Baxter equation, symmetric
functions and Schubert polynomials, to appear in Proceedings of the 5th
International Conference on Formal Power Series and Algebraic Combinatorics,
Firenze, 1993.
\item{[FK2]} S. Fomin, A.N. Kirillov, Universal exponential solutions of the
Yang-Baxter equation, manuscript, 1993.
\item{[FRT]} L.D. Faddeev, N.Yu. Reshetikhin, L.A. Takhtajan, Quantization of
Lie groups and Lie algebras, Algebra and Analysis, 1, 1989, no. 1, 193-225.
\item{[FS]} S. Fomin, R. Stanley, Schubert polynomials and the nilCoxeter
algebra, Advances in Math., to appear; see also Report No. 18 (1991/92),
Institute Mittag-Leffler, 1992.
\item{[Ji]} M. Jimbo, Introduction to the Yang-Baxter equation, Int. Journ.
Mod. Phys. A, 4, 1989, 3759-3777.
\item{[Jo]} V.F.R. Jones, On knot invariants related to statistical models,
Pacific Journ. Math., 137, 1989, 311-334.
\item{[Ko]} A. Kohnert, Weintrauben, Polynome, Tableaux, Thesis, Bayreuth,
1990.
\item{[KB]} A.N. Kirillov, A.D. Berenstein, Groups generated by involutions,
Gelfand-Tsetlin patterns and combinatorics of Yang tableaux, Preprint
RIMS-866, 1992.
\item{[KR]} A.N. Kirillov, N.Yu. Reshetikhin, Representations of the algebra
$U_q(sl(2))$, $q$ - orthogonal polynomials and invariants of links, Adv. Series
in Math. Phys., v. 7, 1989, 285-339.
\item{[KS]} A.N. Kirillov, F.A. Smirnov, The formfactors in the $SU(2)$ -
invariant Thiring model, Journal of Soviet Math., 47, 1989, 2423-2449.
\item{[La]} A. Lascoux, Anneau de Grothendieck de la variete de drapeaux, in
The Grothendieck Festtchrift, vol. III, Birkhauser, 1990, 1-34.
\item{[LS]} A. Lascoux, M.-P. Schutzenberger, Symmetry and Flag manifolds, in
Invariant Theory, Springer lect. Notes in Math., 996, 1983, 118-144.
\item{[Lu]} G. Lusztig, Canonical bases arising from quantized enveloping
algebras, Journ. Amer. Math. Soc., 3, 1990, 447-489.
\item{[M1]} I. Macdonald, Symmetric Functions and Hall Polynomials, Clarendon
Press, Oxford, 1979.
\item{[M2]} I. Macdonald, Note on Schubert polynomials, Laboratoire de
combinatoire et d'informatique mathematique (LACIM), Universite du Quebec a
Montreal, Montreal, 1991.
\item{[MRR]} W.H. Mills, D.P. Robbins, H. Rumsey, Alternating sign matrices
and descending plane partitions, Journ. Combin. Theor., Ser. A, 34, 1983,
340-359.
\item{[R]} J.D. Rogawski, On modules over the Hecke algebras of a $p$ - adic
group, Invent. Math., 79, 1985, 443.
\item{[RT]} N.Yu. Reshetikhin, V. Turaev, Invariants of 3-manifolds via link
polynomials and quantum groups, Inventiones Math., 103, 1991, 547-597.
\item{[S]} R. Stanley, On the number of reduced decompositions of elements of
Coxeter groups, European Journ. Combin., 5, 1984, 359-372.
\item{[T]} T. Tokuyama, A generating function of strict Gelfand patterns and
some formulas on characters of general linear groups, Journ. Math. Soc. Japan,
40, 1988, 671-685.
\item{[V]} D. Verma, Structure of certain induced representations of complex
semisimple Lie algebras, Bull. AMS, 74, 1968, 160-166.
\item{[VK]} N.J. Vilenkin, A.U. Klimyk, Represntation of Lie Groups and Special
Functions, vol 3, Math. and Its Applicat., vol 75, Kluwer Academic Publishers,
1992.
\item{[Y]} C.N. Yang, Some exact results for the many-body problem in one
dimension with repulsive delta-function interaction, Phys. Rev. Lett., 19,
1967, 1312-1314.
\item{[Z]} D. Zeilberger, Proof of the alternating sign matrix conjecture,
Preprint, 1993.

\end